\documentstyle[sprocl]{article}

\def\be{\begin{equation}}
\def\ee{\end{equation}}
\def\bea{\begin{eqnarray}}
\def\eea{\end{eqnarray}}

\begin{document}
\title{The Superhot, Superdense, Supersymmetric Universe~\footnote{Talk given at the SUSY97 Conference, May 27-31, Philadelphia PA, USA.}}
\author{ A. RIOTTO }
\address{NASA/Fermilab Astrophysics Center, \\ Fermilab
National Accelerator Laboratory, Batavia, Illinois~~60510-0500, USA}
\maketitle\abstracts{In this talk the following question is addressed: are    internal symmetries necessarily 
restored at high temperature in  supersymmetric theories? Contrary to the general belief,  we argue that the answer is no when     
systems possess a net background charge. 
Our findings are exemplified  on  abelian models, for both cases of global
and local symmetries and  possible cosmological  implications are also discussed.}

\section{THE PREHISTORIC ERA}
When heated up, physical systems 
undergo phase transitions from 
ordered to less ordered phases. This deep belief, encouraged by 
everyday life experiences, would tell us that at high temperature 
spontaneously broken symmetries of high energy physics get restored.
This, in fact, is what happens in the Standard Model (SM) of electroweak 
interactions. Whether or not true in general is an important question in 
its own right, but it also has a potentially dramatic impact on cosmology.
Namely, most of the extensions of SM tend to suggest the existence of the 
so called topological defects and it is known that two types of such 
defects, {\it i.e.}domain walls and monopoles pose cosmological catastrophe. 
More precisely, they are supposed to be produced during phase 
transitions at high temperature $T$ \cite{kibble} and they simply carry
 too much energy 
density to be in accord with the standard big-bang cosmology.
One possible way out of this problem could be provided by eliminating 
phase transitions if possible. In fact, it has been known for a long time
 \cite{goran} that in theories with a more than one Higgs field (and the
existence of topological defects requires more than one such field in 
realistic theories) symmetries may remain broken at high $T$, and even 
unbroken ones may get broken as the temperature is increased. This offers
a simple way out of the domain wall problem \cite{gia}, whereas the 
situation regarding the monopole problem is somewhat less clear
 \cite{monopole}.

Unfortunately, the same mechanism seems to be inoperative 
in supersymmetric theories. Whereas supersymmetry (SUSY) itself gets broken
 at high $T$, 
internal symmetries on the other hand get necessarily restored. This 
has been proven at the level of renormalizable theories \cite{mangano}. The argument goes as follows: in supersymmetric theories at finite temperature, the leading term in the effective potential is given by $T^2 \left({\rm Tr}{\cal M}^{\dagger}_f{\cal M}_f +{\rm Str}{\cal M}^2\right)$, where ${\cal M}_f$ is the mass matrix of the femionic degrees of freedom and ${\rm Str}{\cal M}^2$ is a field-independent quantity proportional to the soft breaking mass terms. It is easy to show that $\delta {\rm Tr}{\cal M}_F{\cal M}^{\dagger}_f/\delta\varphi_i=0$ gives  $\langle \varphi_i \rangle=0$ \cite{mangano}. 

Recent attempt to evade this argument using higher dimensional 
nonrenormalizable operators \cite{tamvakis}, has been shown not to work
 \cite{borut}.
\section{THE DARK AGES}

About a  year ago,  at the SUSY96 Conference, 
all these considerations led my friends  Borut Bajc and G.  Senjanovi\'c to whisper (or, in other words, with the same tone as when you say  {\it Adelante Pedro, ma con judicio}, A. Manzoni,   {\it I promessi sposi}) that 
 "....
 internal symmetries in supersymmetric theories 
seem to get restored at high temperature, even in the presence of 
non-renormalizable interactions. However, one must admit that the proof 
offered is valid only for a single chiral superfield. We {\it suspect} that 
the above is true in general....". \cite{susy96}. 

As we know, dark ages are necessarily  followed by a delightful  period of renaissance. Luckily enough, that {\it suspect} has been shown recently  to be false. 

\section{ THE RENAISSANCE ERA}

All the considerations mentioned  above  about SUSY and broken symmetries at high
temperatures have an important assumption in common: the 
chemical potential is taken to be zero. In other words, it is assumed 
the vanishing of any  conserved charge.  What is the  fate of internal symmetries at high temperatures
in SUSY theories if  we  relax the assumption of vanishing conserved
charge in the system?  The answer to this question has been recently given in \cite{noi}.  Actually, the answer was  already known in
nonsupersymmetric theories where it has been proven  that the background
charge asymmetry may postpone symmetry restoration 
at high temperature \cite{haberweldon}, and even more remarkably that it can
lead to symmetry breaking of internal symmetries, both in cases of
 global  \cite{scott} 
and local symmetries \cite{linde} at arbitrarily high temperatures. This 
is simply a consequence of the fact that, if the conserved  charge stored
 in the system is larger than a critical value, 
 the charge cannot entirely reside in the thermal excited modes, but it
 must flow into  the 
vacuum.  This is an indication 
that the expectation value of the charged field is non-zero, {\it i.e.} that 
the symmetry is spontaneously broken. 
Furthermore, from the work of Affleck and 
Dine \cite{affleck} we know that there is nothing unnatural about 
large densities in SUSY theories. 

To leave the medieval age and come up to the air, let us warm up with the  simplest supersymmetric
 model with a global $U(1)$ symmetry is provided by a 
chiral superfield $\Phi$ and a superpotential 
\begin{equation}
W = {\lambda\over 3}  \Phi^3.
\label{superpot}
\end{equation}
It has a global $U(1)$ $R$-symmetry under which fields transform as 
\begin{equation} 
\phi \to {\rm e}^{i\alpha} \phi\:\:\:{\rm  and } \:\:\:\psi \to 
{\rm e}^{-i\alpha/2} \psi,
\end{equation}
where $\phi$ and $\psi$ are the scalar and the fermionic component of the 
superfield $\Phi$.
Thus the fermionic and bosonic charges are related by 
$2 Q_\psi + Q_\phi = 0$, or 
in other words the chemical potentials satisfy the relation
\begin{equation}
\mu\equiv   \mu_\phi = - 2 \mu_\psi.
\end{equation}
 The presence of a nonvanishing net
 $R$-charge  leads to a mass term $-\mu^2 \phi^\dagger\phi$ for the scalar  
field   with a ``wrong''
 (negative) sign after canonical momenta   relative to the boson  
degrees of freedom have been integrated out in the
 path integral \cite{haberweldon}. The crucial new ingredient here is that
such a bosonic   
contribution to the mass squared term already appears at the tree-level  
(and therefore is not depending upon $\lambda$), while the $\mu^2$  
dependent part of the fermionic contribution to the mass squared term   
only appears at the one-loop level and is therefore suppressed by  
$\lambda^2$. To see this explicitly, one has to compute
the fermionic-loop tadpole  at finite temperature and  chemical potential.
Performing the summation over the Matsubara modes and  integrating over
the momentum $p$ one may show that 
for a small Yukawa coupling $\lambda\ll 1$ and   $\mu<T$,  the fermionic
contribution in the chemical potential term to the mass  squared is of
order of $\lambda^2 \mu^2 \phi^\dagger\phi$ and therefore  suppressed for
small $\lambda$. We refer the reader to 
ref. \cite{harrington} for more  details. As a result, the fermionic
degrees of freedom cannot compensate the 
 genuine term $-\mu^2 \phi^\dagger\phi$ originated by the bosonic
partners. At finite temperature and finite chemical potential SUSY is
broken. 

The one-loop high temperature
for  small Yukawa coupling $\lambda\ll 1$  reads
 (for $\mu<T$) 
\begin{equation}
V_T(\phi) = \left( -\mu^2  +  {1\over 2} \lambda^2  T^2\right) \phi^\dagger
 \phi + \lambda^2 (\phi^\dagger \phi)^2. 
\label{Tpotential}
\end{equation}
  Obviously, for 
$\mu^2 > \lambda^2 T^2/2$ the symmetry is spontaneously broken at 
high temperature and the field 
$\phi$ gets a vacuum expectation value (VEV)
\begin{equation}
\langle\phi \rangle^2 = {\mu^2 - {\lambda^2\over 2}T^2\over \lambda^2}.
\label{vev}
\end{equation}
This result is valid as long as the chemical potential $\mu$ is smaller 
than the scalar mass in the $\langle\phi\rangle$-background, {\it i.e.}
 $\mu^2 < m_{\phi}^2 = 2 \lambda^2 \langle\phi\rangle^2$ \cite{scott}. 
This in turn implies $\mu > \lambda T$.
In short, for a perfectly reasonable range
\begin{equation}
\lambda T < \mu < T
\label{reasonable}
\end{equation}
the original $U(1)_R$ global symmetry is spontaneously broken 
and this is valid at arbitrarily high temperatures (as long as the 
approximation of $\lambda$ small holds true).  

Notice that, 
in all the above we have assumed unbroken supersymmetry. When supersymmetry
 is softly broken, $U(1)_R$ gets also explicitly broken  because of the
 presence of soft trilinear scalar couplings in the Lagrangian. Therefore,
 the associated net charge vanishes and the reader might be worried 
about the validity of our result. However, the typical rate 
for $U(1)_R$-symmetry 
breaking effects is given by $\Gamma\sim \widetilde{m}^2/T$, where we have 
indicated by $\widetilde{m}\sim 10^2$ GeV the typical soft SUSY 
breaking mass term. Since the expansion rate of the Universe is given
 by $H\sim 30\:T^2/M_{P\ell}$,
$M_{P\ell}$ being the Planck mass,  one finds that 
$U(1)_R$-symmetry breaking effects are in equilibrium and the net charge 
must vanish  only  at  temperatures {\it smaller} than 
   $T_{{\rm SS}}\sim\widetilde{m}^{2/3}M_{P\ell}^{1/3}\sim 10^7$ GeV. 
Therefore, it is perfectly legitimate to consider the presence of a
 nonvanishing $R$-charge at very high temperatures  even in the case of
 softly broken SUSY. 

Thus, we have  provided a simple and natural counterexample to the 
theorem of the restoration of internal symmetries in supersymmetry 
 \cite{noi}. This is a consequence of the fact that the charge cannot
be stored in the 
thermal excited modes, but it must reside in the vacuum and this is an 
indication that the expectation value of the charged field is non-zero, 
{\it i.e.} that the symmetry is spontaneously broken. 

What about gauge symmetries?  It has been known for a long time
\cite{linde} that a background charge asymmetry tends to increase symmetry 
 breaking in the case of a local gauge symmetry in nonsupersymmetric
theories. In his work, Linde has 
shown how a large fermion number density would prevent symmetry restoration
 at high temperature in both 
abelian \cite{linde} and nonabelian theories \cite{linde1}. 
The essential point is that the external charge leads to the condensation
of the gauge field which in turn implies the nonvanishing VEV of the 
Higgs field. This phenomenon may be easily understood if one recalls that 
an increase of an external fermion current ${\bf j}$ leads to symmetry 
restoration in the superconductivity theory \cite{super}. In gauge 
theories symmetry breaking is necessarily a function of $j^2=j_0^2-{\bf j}^2$,
 where $j_0$ is the charge density of fermions. An increase of $j_0$  
is therefore accompanied by an increase of symmetry breaking \cite{linde}. 
We now demonstrate that this phenomenon persists in supersymmetric 
theories, at least in the case of abelian symmetry.

The  simplest model is based on $U(1)$ supersymmetric local gauge symmetry. 
The minimal anomaly free matter content consists of two chiral 
superfields $\Phi^+$ and $\Phi^-$ with opposite gauge charges and the 
 most general renormalizable superpotential takes the form
\begin{equation}
W = m \Phi^+ \Phi^-.
\label{gaugeW}
\end{equation}
Notice that the symmetry is $\it not$ spontaneously broken
at zero temperature.
Since there is no Yukawa interaction, there is also a global 
$U(1)$ $R$-symmetry, under which the bosons have, say, the same charge and
  fermions are invariant. Furthermore at very high temperature, $T>
m^{2/3}M_{P\ell}^{1/3}$,
the fermion mass can be neglected and we get also a chiral $U(1)$ symmetry
 under which the bosons are invariant.

We may now suppose for simplicity that there is a net background charge
 density  $j_0$, with the zero current density 
and that it lies entirely in the fermionic sector. In other words we assume
 the background charge to be in the form of the chiral fermionic charge.
Thus only the fermions have a nonvanishing chemical potential.

Equally important, we assume that the gauge charge of the Universe is zero,
just as in \cite{linde}. In the realistic version of this example, one  
would  imagine the gauge charge to be the electromagnetic one and the
chiral fermionic charge to be, say, the lepton charge in the MSSM. We know from
observation that the electromagnetic charge of the Universe vanishes to 
a good precision. Thus we have to minimize the action with the constraint
that the electric field is zero. What will happen is that some amount of
 bosonic charge will get stored into the vacuum in order to compensate 
for the fermionic one and achieve the vanishing of the electric field.
In this way, the total $U(1)$ charge density of the system including 
the charge of the condensate is equal to zero even if symmetry is broken and
 the gauge forces are short-range ones. 
This is the principal reason behind the resulting
spontaneous symmetry breaking 
of the local gauge symmetry, as we show below. It is crucial thus to have 
some nonvanishing external background charge, {\it i.e.} the model should have 
some extra global symmetry as provided by our chiral symmetry.  

We can obviously take $A_i=0$ in the vacuum and treat $A_0$ on 
the same footing with the scalar fields $\phi^\pm$ (due to the net charge 
$A_0$ cannot vanish in the vacuum). If we now  integrate out $A_0$ using its
equation of motion, assuming the electric field to be zero, we can then 
 compute the high temperature potential for the scalar fields in question 
at high temperature and large charge density  with the following result
 \begin{eqnarray}
V_{\rm eff}(T) &= &{g^2 \over 2} T^2\left(|\phi^+|^2 + |\phi^-|^2\right) \nonumber\\
&+&{g^2 \over 2}  \left(|\phi^+|^2 - |\phi^-|^2\right)^2 \nonumber \\
&+ &  {1 \over 2} {j_0^2 \over 2 \left(|\phi^+|^2 + |\phi^-|^2\right) + T^2},
\label{eff}
\end{eqnarray}
where we have taken $T\gg m$ and we have included both scalar and fermionic
 loop contributions in the $T^2$ mass term for $A_0$. 
Now, except for the $D$-term, the rest of the potential depends only on the 
sum $\phi^2 \equiv |\phi^+|^2 + |\phi^-|^2 $, and thus the energy is 
minimized for the vanishing of the $D$-term potential, {\it i.e.} for 
$|\phi^+|^2 = |\phi^-|^2$. It is easy to see that in this case the 
effective potential has two extrema: 
\begin{equation}
\phi = 0 \quad {\rm and} \quad \phi^2 = {j_0 \over\sqrt{2} g T} 
- {T^2 \over 2}.
\label{vevT}
\end{equation}
The second extremum obviously exists only for 
\begin{equation}
j_0 > {g T^3\over\sqrt{2}}.
\end{equation}
Moreover in that case it is an absolute minimum, while $\phi = 0$ is
a maximum.  
Now,  we can rephrase the above condition in the 
language of the chemical potential (using  $g_* = 4$ )
\begin{equation}
\mu > g T.
\label{mulocal}
\end{equation}
For $ g\ll1$, which is of our interest, $\mu $ easily satisfies 
the condition $\mu < T$. As noted above, since Yukawa interactions are
 absent, the role of the external charge may have been played  by the
 $R$-charge in the scalar sector, the two scalars being equally charged
 under this symmetry. In such a case, the analysis requires  careful 
handling because of issues related to gauge invariance.   
In a future publication \cite{future} we will extensively  explore this issue as weel generalize our results to theories containing an arbitrary number of 
 abelian symmetries and to nonabelian theories. 

\section{THE FUTURE DAYS} 

We hope to have convinced the audience that 
 internal symmetries in supersymmetric theories,
 contrary to the general belief, may be broken at high temperature, as
long as the system has a nonvanishing background charge. The examples 
we have provided here, based on both global and local abelian symmetries,
are natural and simple and should be viewed as prototypes of more
realistic theories. The necessary requirement for the phenomenon
to take place is that the chemical potential be bigger than a fraction
of temperature on the order of (1-10)\%. Notice that this is by no 
means unnatural. In the expanding universe 
the chemical potential is proportional to temperature, and thus unless
zero for some reason, $\mu /T$ is naturally expected to be of order
one. More important, this chemical potential could be zero today, all
that is needed is that it is nonvanishing at high temperature. We 
have seen how soft supersymmetry breaking may naturally provide such a 
scenario if there is some nonvanishing external charge.
 
 Now, it is well known that in suspersymmetry the existence of flat 
directions may lead to large baryon and lepton number densities at very
high temperature \cite{affleck}. The most natural candidate for the large density of the universe is the
lepton number that may reside in the form of neutrinos.  It should be stressed that a large lepton number is perfectly
consistent with the ideas of grand unification. It can be shown that in
$SO(10)$ one can naturally arrive at a small baryon number and a large
lepton number \cite{hk81}.

In any case, we wish to be even more
open minded and simply allow for a charge density without worrying about 
its origin. It is noteworthy that a large neutrino number density
may persist all the way through nucleosynthesis up to today 
\cite{chemicalpotential}. This 
has been used by Linde  \cite{linde1} in order to argue that even in SM
the $SU(2)_L\otimes U(1)_Y$ symmetry may not be restored at high temperature. 
Since SUSY, as we have seen, does not
spoil the possibility of large chemical potentials allowing symmetry 
breaking at high $T$, it is important to see if our results remain 
valid in the case of nonabelian global and local symmetries. This work
 requires particular attention due to the issues related to gauge invariance
 and is now in progress \cite{future}.    

We should stress that there is more than a sole academic interest
to the issue discussed in this paper. If symmetries remain broken at 
high temperature,
there may be no domain wall and monopole problems at all. Furthermore,
it is well known that in SUSY  grand unified theories symmetry restoration
 at high temperature prevents the system from  leaving the false vacuum and 
finding itself  in the broken phase at low $T$. This is a direct consequence 
of the vacuum degeneracy characteristic of supersymmetry which says that at
 zero temperature the SM vacuum and the unbroken GUT symmetry one have the same
(zero) energy.
If the symmetry is restored at high $T$, one would start with the unbroken
 symmetry in the early universe and would thus get caught in this state 
forever. Obviously, if our ideas 
hold true in realistic grand unified theories, this problem would not
arise in the first place. Moreover, we know that the presence of anomalous baryon number violating precesses induced by sphaleron transitions pose a serious problem for the survival of the    baryon asymmetry in the early Universe. However, sphalerons are active at high temperatures only if the standard model gauge symmetry is restored. The presence of a nonvanishing (lepton?) conserved charge in the minimal supersymmetric standard model might cause the symmetry nonrestoration of the gauge symmetry when the system is heated up, as originally suggested by Liu and  Segr\`e for the nonsupersymmetric standard model \cite{ls94}. This would  allow to freeze the baryon number violating processes and  to 
preserve the baryon asymmetry generated at the GUT scale.

\section*{Acknowledgments} It is with immense  pleasure that we  thank my collaborators  Borut  Bajc and Goran  Senjanovi\'c with whom this journey out of the dark ages towards a so exciting   period has begun. 
 This work is supported by the DOE and NASA
under Grant NAG5--2788.

\end{document}